\documentclass[english,reprint]{revtex4-2}
\usepackage{lmodern}

\usepackage[T1]{fontenc}
\usepackage[latin9]{inputenc}
\usepackage{babel}
\usepackage{amsmath}
\usepackage{graphicx}
\usepackage[unicode=true,pdfusetitle,
 bookmarks=true,bookmarksnumbered=false,bookmarksopen=false,
 breaklinks=false,pdfborder={0 0 1},backref=false,colorlinks=false]
 {hyperref}
\hypersetup{
 linkcolor=blue, urlcolor=blue, citecolor=blue}
\begin{document}
\title{Local and extensive fluctuations in sparsely-interacting ecological
communities}
\author{Stav Marcus, Ari M Turner and Guy Bunin}
\affiliation{Department of Physics, Technion - Israel Institute of Technology,
Haifa 32000, Israel}
\begin{abstract}
Ecological communities with many species can be classified into dynamical
phases. In systems with all-to-all interactions, a phase where a fixed
point is always reached and a dynamically-fluctuating phase have been
found. The dynamics when interactions are sparse, with each species
interacting with only several others, has remained largely unexplored.
Here we show that a new type of phase appears in the phase diagram,
where for the same control parameters different communities may reach
either a fixed point or a state where the abundances of a finite subset
of species fluctuate, and calculate the probability for each outcome.
These fluctuating species are organized around short cycles in the
interaction graph, and their abundances undergo large non-linear fluctuations.
We characterize the approach from this phase to a phase with extensively
many fluctuating species, and show that the probability of fluctuations
grows continuously to one as the transition is approached, and that
the number of fluctuating species diverges. This is qualitatively
distinct from the transition to extensive fluctuations coming from
a fixed point phase, which is marked by a loss of linear stability.
The differences are traced back to the emergent binary character of
the dynamics when far away from short cycles in the local fluctuations
phase.
\end{abstract}
\maketitle
Ecosystems can be extremely diverse \citep{may_how_1988}, and the
large numbers of species makes statistical mechanics a powerful tool
in addressing such systems. Theoretical and experimental work has
revealed dynamical phases with distinct behavior \citep{rieger_solvable_1989,opper_phase_1992,galla_dynamics_2005,kessler_generalized_2015,bunin_ecological_2017,tikhonov_collective_2017,altieri_properties_2021,bunin_directionality_2021,marcus_local_2022,hu_emergent_2022},
with a notable example being a transition from a phase where dynamics
reach fixed points to one where variables fluctuate indefinitely,
which is marked by a loss in the fixed points' stability \citep{rieger_solvable_1989,opper_phase_1992,galla_dynamics_2005,bunin_ecological_2017,hu_emergent_2022}.
Related transitions are also found in other fields \citep{sompolinsky_chaos_1988,galla_complex_2013,fyodorov_nonlinear_2016,guo_exploring_2021}.

Most research in the field has assumed that each species interacts
with many others. Much less is known about sparsely-interacting systems,
where each species interacts with only a handful of the (many) other
species, despite evidence that such systems are ubiquitous in nature
\citep{dunne_food-web_2002,busiello_explorability_2017}. Previous
works on sparse interactions have generalized bounds on diversity
from fixed-point stability \citep{may_will_1972} to sparse settings
\citep{allesina_stability_2012,mambuca_dynamical_2022}, and obtained
statistics of the number of fixed points in the limit of strong interactions
\citep{fried_alternative_2017}. For symmetric interactions, properties
of fixed points \citep{marcus_local_2022} and activated dynamics
\citep{bunin_directionality_2021} have been found to differ dramatically
from their fully-connected counterparts. This work will focus on autonomous
persistent dynamics, not found for symmetric interactions which always
reach fixed points in the absence of external noise \citep{macarthur_species_1970,pykh_lyapunov_2001}.

\begin{figure*}
\begin{centering}
\includegraphics[viewport=0bp 5bp 1904bp 620bp,clip,width=1\textwidth]{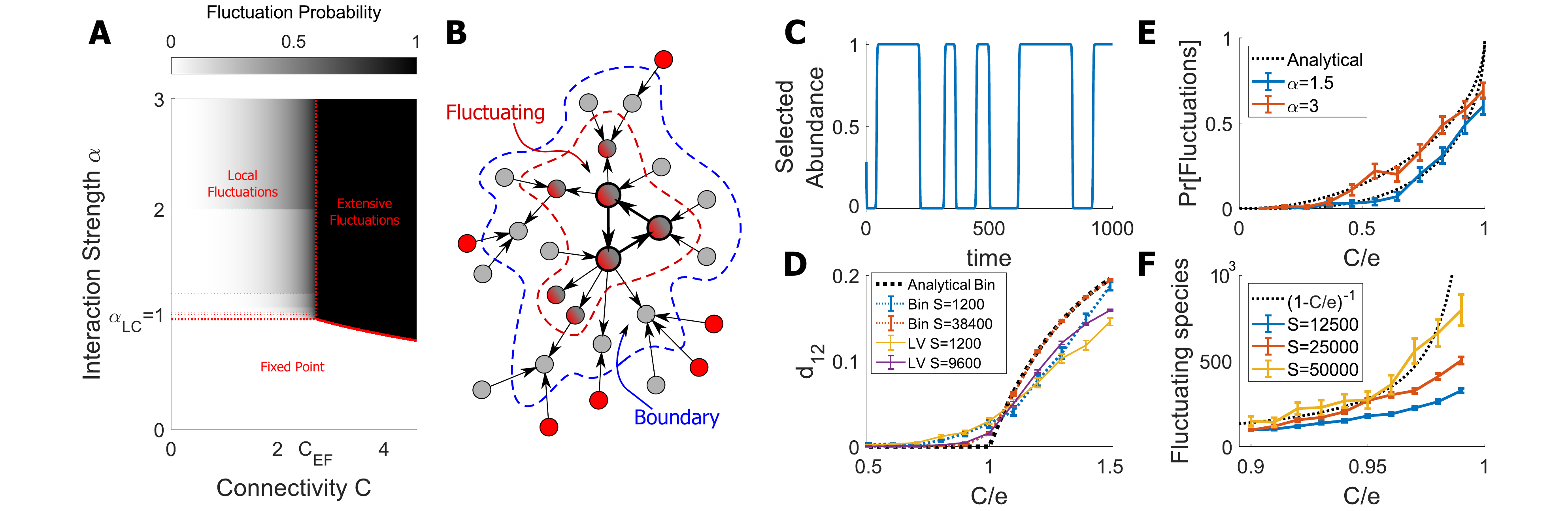}
\par\end{centering}
\caption{\label{fig:Simplified-model}\textbf{The minimal model.} \textbf{(A)
}Phase diagram and fluctuation probability in the space of connectivity
$C$ and interaction strength $\alpha$. Thick red lines mark transition
lines between the phases, and thin lines in the local fluctuation
phase mark discontinuous jumps in fluctuation probability, at the
values $\alpha_{c}(n)$. \textbf{(B)} A short cycle that might fluctuate
in the LF phase, and its neighborhood. The fluctuations originate
on a cycle (with thicker arrows), here of length three, spreading
downstream until reaching either an extinct species or one that affects
no other species. All boundary species to this fluctuating subset
are extinct, allowing the cycle to fluctuate and isolating this subset
from the rest of the system.\textbf{ }The length-3 cycle then fluctuates
if $\alpha>\alpha_{c}(3)=2$. Fluctuations are then as shown, with
$N_{i}=1$ species in red, $N_{i}=0$ in gray and fluctuating $N_{i}$
in red-gray. \textbf{(C)} Example of abundance dynamics at $\alpha>1$
far away from any short cycles. Variables spend most of the time near
$N_{i}=0,1$, and may alternate between them with rapid transitions.
Taken from system with $S=1000,C=4,\alpha=3$.\textbf{ (D) }The distance
$d_{12}$ between replicas after a long time as a function of $C$.
It continuously increases from zero at $C>e$ in both the LV model
(here with $\alpha=3$) and the binary model. \textbf{(E) }The probability
of fluctuations as a function of $C$, comparing analytical result
(dashed lines) with simulations at $S=1000$ and for $\alpha=1.5<\alpha_{c}^{(3)}$,
$\alpha=3>\alpha_{c}^{(3)}$. The probability reaches $1$ at the
transition to the EF phase. \textbf{(F) }The number of fluctuating
species as a function of $C$, from simulations of the binary model
at different system sizes $S$. It converges to the expected behavior
of $\left(1-C/e\right)^{-1}$ near the transition, the trend shown
in a dashed black line (with prefactor chosen to fit the data).}
\end{figure*}

Here, we study the dynamics of sparsely-interacting communities of
species when interactions are far from symmetric. We first study a
minimal model where interactions are unidirectional (i.e. one species
affects another but not vice versa), then extend the analysis. Our
analysis reveals three phases, see Fig. \ref{fig:Simplified-model}(A):
(1) A fixed point (FP) phase, where dynamics always reach a stable
fixed point; (2) A local fluctuations (LF) phase, where the abundances
of a small number of clustered species fluctuate, while those of the
rest are fixed. (3) An extensive fluctuations (EF) phase, where the
abundances of a finite fraction of the species fluctuate. The LF phase
arises in the sparse setting, and cannot appear when there are all-to-all
interactions.

The fluctuations in the LF phase are organized around fluctuating
centers, each involving a finite set of species. We study these centers
and their statistical properties when embedded within the large network.
We calculate the probability that a system will reach a dynamical
state that fluctuates indefinitely, a probability that is experimentally
accessible and of great interest \citep{hu_emergent_2022}. We find
that in the LF phase this probability remains between zero and one,
even for large systems; this is in contrast to fully-connected networks,
where it is either zero (at a fixed-point phase) or one (in a fluctuating
phase). When changing the control parameters to approach the transition
to the EF phase, this probability continuously increases to one and
remains at one in the EF phase.

We then compare the transition to the EF phase when coming from either
the LF or the FP phase. We see that the latter is accompanied by a
divergence of a response function, similar to a divergence found in
models with all-to-all interactions. In contrast, this response function
\emph{does not diverge} when transitioning from the LF phase to the
EF phase, showing a qualitative difference between the transitions.

Last, we show that the distinctions between the phases and the respective
transitions extend to more generic systems beyond the minimal model,
that may include variability in interactions strength and bi-directionality
of interactions, through multiple properties of these phases.

The existence of localized dynamics that we find may be relevant to
situations in nature where only a few species fluctuate \citep{elton_periodic_1924,utida_population_1957,peterson_wolf_1999,beninca_species_2015},
often modeled by few variable dynamics \citep{lotka_elements_1925,volterra_fluctuations_1926,nicholson_balance_1935,may_nonlinear_1975,park_evolutionary_2021,beninca_species_2015}.
Our work shows how and when localized dynamics might occur even for
species that form part of a large ecological community, as almost
always happens in nature \citep{stenseth_population_1997}.

We employ the standard Lotka-Volterra (LV) dynamics for a community
assembled from a pool of $S$ species. The dynamics of the abundance
$N_{i}$ of species $i$ in the community is given by

\begin{equation}
\frac{dN_{i}}{dt}=N_{i}\left(K_{i}-\sum_{j=1}^{S}\alpha_{ij}N_{j}\right)+\lambda_{i}\ .\label{eq:LV}
\end{equation}
The abundances are rescaled so that the carrying capacities are $K_{i}=1$,
and the pool migration rates are taken to be small, $\lambda_{i}\ll1$.
We start with a simple version of the model, which nevertheless exhibits
rich phenomenology, and is later extended. Here, the intraspecific
interaction for all species is $\alpha_{ii}=1$, and all other interaction
strengths $\alpha_{ij}$ are drawn independently, with $\alpha_{ij}=\alpha$
with probability $C/S$, and $\alpha_{ij}=0$ otherwise. Graphically,
this forms a directed Erd\H{o}s-Rényi graph, with an edge from $i$
to $j$ if $\alpha_{ji}=\alpha$. This graph is sparse, with an average
of $C$ incoming and $C$ outgoing edges per species, even at large
$S$. $C$ is the called the connectivity. Some known facts regarding
these graphs are: almost all edges $i\rightarrow j$ do not have a
reverse $j\rightarrow i$; the local neighborhood of almost all vertices
is a tree (no cycles); there are only a few cycles of any finite length
(their average number doesn't scale with $S$); and finite-length
cycles are isolated (the distance between them grows with $S$) \citep{bollobas_random_2001}.
Unidirectional interactions are taken as a starting point as they
are very different from symmetric interactions where dynamics always
reach fixed points \citep{macarthur_species_1970}.

\emph{Onset of extensive fluctuations}--When $\alpha>1$, we see
two distinct dynamical behaviors depending on $C$, with a transition
at $C_{\text{EF}}$which we show below to be $C_{\text{EF}}=e$. For
$C<C_{\text{EF}}$, all species far from short cycles are fixed at
either $N_{i}\approx1$ or $N_{i}\approx0$ (both to order $\lambda_{i}$,
henceforth we write $N_{i}=0,1$). At $C>C_{\text{EF}}$, an extensive
number of species fluctuate. The fluctuating species spend most of
the time near $0,1$ with rapid transitions between them, see Fig.
\ref{fig:Simplified-model}(C). The timescale between transitions
is $O\left(\left|\ln\lambda\right|\right)$, corresponding to the
time for a species $N_{i}$ to grow exponentially from $O\left(\lambda\right)$
to order one, while the time of the transitions themselves doesn't
depend on $\lambda$.

These rapid transitions can be captured by a model with binary variables,
$N_{i}\in\left\{ 0,1\right\} $, which will be helpful in deriving
the location of the transition and other properties. In the dynamics
of Eq. (\ref{eq:LV}), flips are determined by history-dependent function
of the inputs to $i$. For some purposes (see below), the exact dynamics
can be replaced by a simple stochastic dynamics where species $i$
transitions to $N_{i}=0$ if at least one of its incoming arrows have
$N_{j}=1$, and to $N_{i}=1$ otherwise, reflecting the growth or
decline of $N_{i}$ with such inputs. The flip is done at some constant
rate, so the only free parameter is the connectivity $C$.

The following extensive properties can be calculated for the binary
model. First, consider the fraction at long times of $N_{i}=1$ species,
$\phi=\Pr[N_{i}=1]$. Since almost all cycles are long, $N_{j}$ can
be taken to be independent between species $j$ incoming to the same
$i$. Using the requirement that after a flip, $N_{i}=1$ if and only
if all $N_{j}=0$, one obtains the self-consistent $\phi=\sum_{K}P_{K}(1-\phi)^{K}$,
where $P_{K}=e^{-C}C^{K}/K!$ is the probability for $K$ incoming
edges in the graph. This is solved by $\phi=W(C)/C$, where $W$ is
the Lambert W function. This is the same value of $\phi$ found for
fixed points of the system in the limit of infinitely strong interactions
by other methods \citep{fried_alternative_2017}. The transition to
the EF phase, marked by the appearance of persistent fluctuations,
can be located using the technique of \citep{derrida_random_1986,derrida_dynamical_1987,derrida_exactly_1987}
as reviewed in the SM. This involves writing a closed equation (depending
also on the known $\phi$) for the distance $d_{12}(t)=\frac{1}{S}\sum_{i=1}^{S}\left|N_{i}^{1}(t)-N_{i}^{2}(t)\right|$
between two copies of the same system with different initial conditions
$\left\{ N_{i}^{1}\right\} ,\left\{ N_{i}^{2}\right\} $. The distance
at long times is zero in the LF phase, and grows continuously from
zero when increasing $C$ above the transition point $C=e$ for both
the binary and LV models (See Fig. \ref{fig:Simplified-model}(D)).

From this we conclude that in the LV model at $\alpha>1$ there is
a transition at $C\equiv C_{\text{EF}}=e$ into an EF phase, see Fig.
\ref{fig:Simplified-model}(D). The above analysis for the binary
model also applies to the LV model: The argument for calculating $\phi$
is valid at a fixed point of the LV model (far from the rare short
cycles). And, if a fixed point is stable in the binary model, it is
also stable in the LV model.

\emph{Local fluctuations--}We now turn to analyze local fluctuations.
As follows from the discussion above, for $C<C_{EF}$ and $\alpha>1$
the variables $N_{i}$ reach a fixed value, except possibly a sub-extensive
fraction. We now characterize the structure of the fluctuating pieces,
show that they are finite, and calculate the probability that the
entire system will reach a fixed point at long times.

Consider a local set of fluctuating species. First, we argue that
the fluctuating species are surrounded by extinct species, and so
effectively isolated from the rest of the community. To do that, look
at the set's boundaries, i.e. species that interact with the fluctuating
species but do not themselves fluctuate, see Fig. \ref{fig:Simplified-model}(B).
If a boundary species $N_{j}$ is upstream from a fluctuating species
$i$, it must be extinct ($N_{j}=0$), otherwise it would not allow
$N_{i}$ to fluctuate: if $N_{j}=1$ and $\alpha_{ij}>1$, $N_{i}$
would be prevented from growing. If this boundary species is downstream
from the fluctuating species it must also be extinct, otherwise it
too would fluctuate. Second, as the dynamics on any finite tree always
reach a fixed point \citep{pykh_lyapunov_2001}, the fluctuating subset
must include a single short cycle (and no more than one, as finite-length
cycles are rare and distant from each other), see example in \ref{fig:Simplified-model}(B).
The fluctuating subset will then include the tree and possibly downstream
species: species upstream from the cycle are unaffected by it, and
so cannot fluctuate since they are effectively a separate tree.

Thus, all incoming species to the cycle are extinct, so the cycle
behaves as if it is isolated. In isolation, it can fluctuate only
if it is directed, i.e. all interactions are directed in the same
sense along the cycle (otherwise the cycle can be broken up into chains
which must reach a fixed point), and if it is of odd length (otherwise
there is a stable fixed point where the species along the cycle alternate
between $N_{i}=0$ and $N_{i}=1$). On these cycles, there exists
a fixed point where the $N_{i}$'s are not $0,1$, but instead $N_{i}=1/(1+\alpha)$
for all the cycle species. There are fluctuations if this fixed point
is unstable, with stability lost for $\alpha>\alpha_{c}(n)=1/\cos(\pi/n)$
for cycles of odd length $n$, see SM. So the cycle would fluctuate
if $\alpha>\alpha_{c}(n)$, and with it the local variables downstream
from it. For example, in the setting shown in Fig. \ref{fig:Simplified-model}(B),
this occurs for $\alpha>\alpha_{c}(3)=2$. For $\alpha<\alpha_{c}(n)$,
a set of species connected in the same way would reach a fixed point,
and the abundances downstream from the cycle and inside the local
configuration are also not $0,1$ and are set by Eq. (\ref{eq:LV}).

Now let us determine the probability of fluctuations. As described
above, fluctuations require three conditions: an odd directed cycle
exists in the graph; it is unstable (this depends on $\alpha$); and
incoming interactions from boundary species do not drive cycle species
to extinction. The probability for these three conditions to be satisfied
can be calculated exactly, as follows: The third condition requires
that all the species upstream of any species in the cycle have abundances
of zero. Since there are no other short cycles in the cycle's vicinity,
the incoming variables are independent from each other (as happens
in a typical tree-like environment, see above). The probability that
none of the incoming species will have $N_{j}=1$, and therefore prevent
$N_{i}$ from fluctuating, is the same as the probability for any
species in the graph not to be driven to extinction, which is exactly
$\phi$. Thus, given a directed cycle of length $n$ that fluctuates
in isolation, the probability that it will fluctuate in the graph
is $\phi^{n}$. The number of directed cycles of length $n$ is Poisson
distributed with mean $C^{n}/n$ \citep{bollobas_random_2001}. Each
fluctuates with probability $\phi^{n}$, so the number of fluctuating
cycles of length $n$ is Poisson distributed with mean $m_{n}=(C\phi)^{n}/n$.
The numbers of cycles of different lengths are independent, so the
total number of cycles is Poisson distributed, and the probability
of at least one cycle fluctuating is $1-e^{-m}$, where $m\equiv\sum_{j=n}^{\infty}(C\phi)^{2j+1}/(2j+1)$
in the range $\alpha_{c}(2n+1)<\alpha<\alpha_{c}(2n-1)$. The calculated
probabilities are shown in Fig. \ref{fig:Simplified-model}(A,E);
note the jumps in probability at the $\alpha_{c}(n)$ where different
cycles lose stability, and the continuous increase with $C$ until
reaching $1$ at \textbf{$C_{EF}$} for all $\alpha$. For $\alpha<1$
all cycles are stable, and the system enters the fixed point phase.

We turn to the downstream effect of these fluctuations. The species
on a fluctuating cycle may drive downstream species to fluctuate as
well, and the fluctuations spread in this way until they are ``blocked'',
when a fluctuating species only affects extinct species (if any).
This downstream spread can be considered as a branching process. The
number of species driven to fluctuate by any fluctuating species is
Poisson distributed with mean $\rho(C)$. As such species are driven
to fluctuate iff they are not extinct, this occurs with probability
$\phi$; and as the average number of outgoing edges is $C$, this
gives $\rho(C)=\phi C=W(C)$. The downstream species from each species
on the cycle (excluding its neighbor along the cycle) form a tree
of fluctuating species. This tree grows with $C$ and becomes infinite,
i.e., the fluctuations percolate, when on average each species drives
one other species to fluctuate, $\rho\left(C\right)=1$. This occurs
exactly at the transition to the EF phase, $C=C_{\text{EF}}=e$. The
number of fluctuating species tends to infinity near the transition
as $S_{\mathrm{fluct}}\sim\left(1-C/C_{\text{EF}}\right)^{-1}$ \citep{harris_theory_1963},
as shown in Fig. \ref{fig:Simplified-model}(E). Importantly, this
means that this number is either $O(1)$ for $C<C_{\text{EF}}$ or
$O(S)$ for $C>C_{\text{EF}}$, with no other behaviors in between.
Above $C_{\text{EF}}$ the abundances of a finite fraction of the
species fluctuate with probability $1$, so the dynamics of small
subgraphs can no longer be understood in isolation. A similar divergence
in the number of fluctuating variables, of a percolation universality
class, occurs in the ``frozen phase'' of binary models such as the
Kauffman model \citep{bastolla_relevant_1998}. In this context, the
fluctuation probability discussed above has not been calculated, and
may be of interest.

To summarize the transition between the LF and EF phases: in the LF
phase, a hybrid binary-continuous behavior emerges, with the variables
taking two values, except close to short cycles that might stabilize
at $N_{i}\notin\{0,1\}$ and be effectively removed from the binary
model. The transition to the EF phase is characterized by three effects:
the probability of fluctuations, which is between zero and one, goes
to one; the number of fluctuating species diverges; and extensive
fluctuations begin. Approaching the FP phase from the LF phase by
lowering $\alpha$, cycles of ever-longer length become stable (at
$N_{i}\notin\{0,1\}$). This reduces the probability of fluctuations
originating from finite-length cycles, until the entire system stabilizes
when crossing the $\alpha=1$ line.

\begin{figure}[!t]
\begin{centering}
\includegraphics[viewport=0bp 0bp 750bp 460bp,clip,width=1\columnwidth]{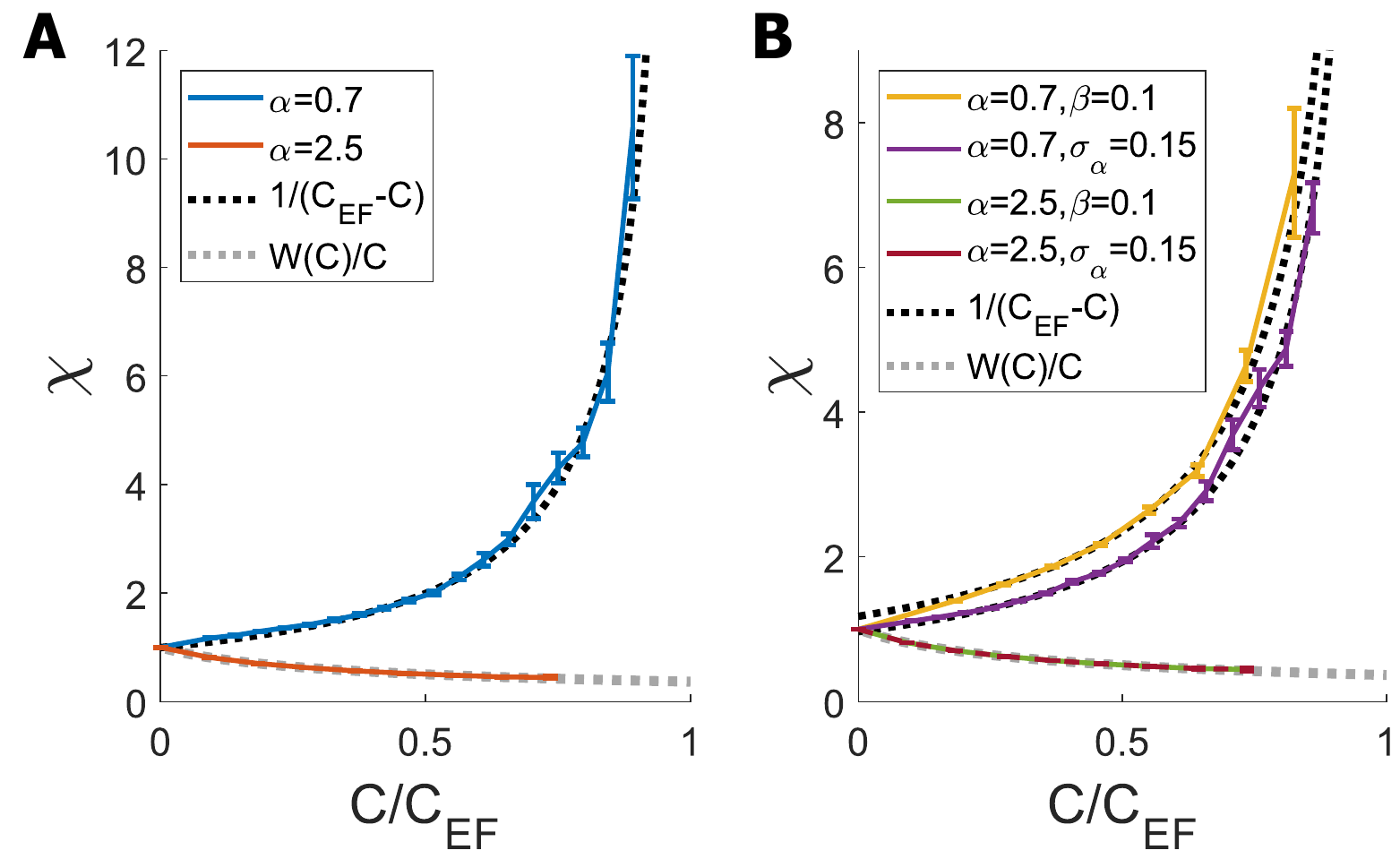}
\par\end{centering}
\caption{\label{fig:chi}\textbf{The response function $\chi$ when approaching
the transition line to the EF phase by increasing the connectivity
$C$}. \textbf{(A)} $\chi$ diverges when increasing $C$ at constant
$\alpha$ coming from the FP phase, indicating the loss of stability
of the fixed point (here $\alpha=0.7$). In contrast, $\chi$ does
not diverge when coming from the LF phase (here $\alpha=2.5$). The
dotted black line is a fit of $A\left(C_{\text{EF}}-C\right)^{-1}$,
and the dotted gray line is the analytical result $W(C)/C$. $C_{\text{EF}}=e$
for $\alpha=2.5$. $C_{\text{EF}}\approx5.3$ from fit to points at
$C>2$ for $\alpha=0.7$. \textbf{(B)} Adding variability and bi-directionality
to the interactions does not change the transition properties: for
$\alpha=0.7$, $\chi$ still diverges as $\left(C_{\text{EF}}-C\right)^{-1}$,
and for $\alpha=2.5$, $\chi=W\left(C\right)/C$ still holds. For
both $\alpha=0.7$ and $\alpha=2.5$, we introduce bi-directionality
by setting $\alpha_{ji}=\beta=0.1$ when $\alpha_{ij}=\alpha$, or
add variability by drawing the non-zero $\alpha_{ij}$ from a normal
distribution $\mathcal{N}(\alpha,0.15)$. For $\alpha=0.7$, $C_{\text{EF}}\approx4.9,2.7$
for $\sigma_{\alpha}=0.15$ and $\beta=0.1$ respectively.}
\end{figure}

\emph{Transition from fixed point to extensive fluctuations}--Besides
the possibility of local fluctuations, there are other distinctions
between the FP and LF phases. Consider the transitions between these
phases and the EF phase. A transition from a fixed-point phase to
an EF phase is found in systems with all-to-all interactions. There,
it is known that the approach to the transition is marked by a divergence
of a response $\chi$, defined as follows. Consider a small change
in the carrying capacities $K_{i}\to1+\varepsilon_{i}$, with $\varepsilon_{i}$
independent random numbers, and let $n_{i}$ be the corresponding
change in the fixed point value of $N_{i}$. Then $\chi\equiv\left\langle n_{i}^{2}\right\rangle /\left\langle \varepsilon_{i}^{2}\right\rangle $
diverges when approaching the transition \citep{opper_phase_1992,bunin_ecological_2017,roy_numerical_2019}.
This corresponds to closing the gap in the spectrum of the matrix
$\left\{ \alpha_{ij}\right\} $ (the real part of the eigenvalues
reaching zero). We find that the same quantity diverges also in sparse
communities when approaching the transition line $C_{\text{EF}}(\alpha)$
from the FP phase, as $\left[C_{\text{EF}}(\alpha)-C\right]^{-1}$,
see Fig. \ref{fig:chi}. This is very different from what happens
in the transition from the local fluctuations phase: there, the species
that don't fluctuate and have $N_{i}=1$ are isolated (surrounded
by extinct species, if any), so $n_{i}=\varepsilon_{i}$, while for
the species with $N_{i}=0$, $n_{i}=0$ . Restricting to the species
that don't fluctuate, which are all but possibly a finite subset,
we therefore have $\chi=\phi$. This quantity was calculated above,
and does not diverge at the transition, see Fig. \ref{fig:chi}. The
subset of fluctuating species does not change due to the application
of a small perturbation. One can extend the definition of $\chi$
to include the fluctuating species, by measuring the change in the
time average of $N_{i}(t)$ due to the changes in $K_{i}$. This response
is also finite, and because it only applies to a finite number of
variables it does not change $\chi$ (Note that the limit $S\to\infty$
is taken first, before varying the system parameters to approach the
transition.). Thus, $\chi=\phi$ in this extended definition as well,
and remains finite when approaching the transition from the LF phase.

\emph{Distinct properties of the FP and LF phases beyond the minimal
model}--The minimal model discussed so far assumes unidirectional
interactions of constant strength $\alpha$. We conclude by discussing
the robustness of the results, by introducing finite but not too large
changes to interaction coefficients; changes beyond some size may
drive the system to other phases not discussed here, such as those
found for symmetric interactions \citep{marcus_local_2022,bunin_directionality_2021}.
The main qualitative results on the transitions to the EF phase are
robust when introducing some variability in $\alpha_{ij}$, and also
bi-directionality, namely allowing for $\alpha_{ji}\ne0$ with a finite
probability when $\alpha_{ij}$ is non-zero.

At low values of $\alpha$, increasing the connectivity $C$ towards
the EF phase is still accompanied by a divergence in $\chi$, even
with the changes to the model, see Fig. \ref{fig:chi}(B). This is
because changes in $\alpha_{ij}$ change the spectrum of the matrix
$\left\{ \alpha_{ij}\right\} $ continuously, so the spectral gap
is not prevented from closing. At large values of $\alpha$, in the
LF phase, the fluctuating subsets (when they exist) remain of finite
size. This is because their downstream extension is limited by the
chance of reaching an extinct species (that can't fluctuate), and
this probability depends continuously on the $\alpha_{ij}$ values;
while the upstream species from fluctuating species are extinct in
the minimal model, and so adding some level of bi-directionality would
not affect them. Furthermore, when adding variability in $\alpha_{ij}$
and bi-directionality around large enough values of $\alpha$, the
neighbors of any species with $N_{i}>0$ that is far away from cycles
are extinct, just as in the minimal model. We prove this for the case
where all $\alpha_{ij}>1$, and the reverse $\alpha_{ji}>0$, see
SM. This means that $\chi$ remains finite when approaching the transition
just as in the minimal model, see Fig. \ref{fig:chi}(B). That the
two qualitatively different scenarios for approaching the EF phase
are robust to these changes in the model (with $\chi$ diverging in
one and not in the other) means that there must be a sharp transition
between these behaviors, an interesting direction for further research.

With variability in $\alpha_{ij}$, the jumps in the probability of
local fluctuations when changing $\alpha$ are broadened. This means
that the sharp boundary between the FP phase and LF phase in terms
of probability of fluctuations is no longer sharp. A sharp distinction
between these phases remains however in the decay of fluctuations,
away from a fluctuating center or a local perturbation. In what is
the FP phase in the minimal model, the size of the fluctuations decays
exponentially with the distance from the source of the fluctuations
but extends arbitrarily far, while in the LF phase it is cut off at
a finite distance. This distinction separates the two phases in the
presence of both variability in $\alpha_{ij}$ and bi-directionality.

\emph{Acknowledgments}--G.B. acknowledges support from the Israel
Science Foundation (ISF) Grant No. 773/18. A.M.T. acknowledges support
from the Israeli Science Foundation (ISF) Grant No. 1939/18.

\bibliographystyle{unsrt}
\bibliography{Local_and_global_fluctuations}

\clearpage

\appendix
\onecolumngrid

\section*{Supplementary material}

\section{\label{sec:critical alpha appendix}Fluctuations in short cycles}

Here we prove that for directed cycles of odd length $n$, fluctuations
occur for $\alpha>\alpha_{\text{cycle}}^{\left(n\right)}=1/\cos\left(\pi/n\right)$,
by showing that they have no stable fixed point in this range. For
a cycle of length $n$, we label the species as $N_{1},...,N_{n}$,
with species $i-1$ affecting species $i$. There are two possible
kinds of fixed points, either with all species having $N_{i}=1/(1+\alpha)>0$
or with some having $N_{i}=0$. If all $N_{i}>0$, the fixed point
is stable if the $\alpha_{ij}$ matrix is stable. Using the eigenvalues
for an $n\times n$ circulant matrix \citep{davis_circulant_1986},
one finds that this fixed point is stable for odd $n$ for $\alpha<1/\cos\left(\pi/n\right)$.
We will now show that there can be no other stable fixed point to
the system, so for $\alpha>\alpha_{\text{cycle}}^{\left(n\right)}$
there will be fluctuations.

Assume by contradiction that the cycle has a stable fixed point with
extinct species. As $n$ is odd, it must have at least two consecutive
species that both have either $N_{i}=0$ or $N_{i}>0$. If there is
a set of consecutive species with $N_{i}>0$, denote as $j$ the first
species in this chain, so that $N_{j}=1$. Therefore $N_{j+1}=1-\alpha N_{j}=1-\alpha<0$,
in contradiction to the assumption. If there are consecutive species
with $N_{i}=0$, again denote $j$ as the first in the chain. Then
species $j+1$ has a positive growth rate, $g_{j+1}=1-\alpha N_{j}=1$,
so the fixed point is not stable to small positive perturbations in
$N_{j+1},$ again in contradiction to the assumption.

\section{\label{sec:DMFT appendix}Distance in the binary model}

In a binary model, at each time step $dt=1/S$ a random species is
chosen to be updated according to
\begin{equation}
N_{i}\left(t+dt\right)=\begin{cases}
1 & \text{all incoming arrows }j\to i\text{ have }N_{j}=0\\
0 & \text{Otherwise}
\end{cases}
\end{equation}
where for two copies of the same system, $\left\{ N_{i}^{1}\right\} ,\left\{ N_{i}^{2}\right\} $,
the same species is updated at each step. Defining the per-species
distance $d_{12,i}=P\left[N_{i}^{1}=N_{i}^{2}\right]$, the total
distance is $d_{12}=S^{-1}\sum_{i}d_{12,i}$. Say that species $N_{i}$
which has $K$ incoming interactions from $N_{j_{1}},..,N_{j_{K}}$is
updated at time $t$. Then as $N_{j_{1}},..,N_{j_{K}}$ are independent,
and using $\Pr\left[\left(N_{i}^{1},N_{i}^{2}\right)=\left(0,0\right)\right]=1-\phi-d_{12,i}/2$
\begin{align*}
d_{12,i}\left(t+dt\right) & =2\Pr\left[N_{i}^{1}\left(t+dt\right)=1,N_{i}^{2}\left(t+dt\right)=0\right]\\
 & =2\left\{ \prod_{r=1}^{K}\Pr\left[N_{j_{r}}^{1}=0\right]-\prod_{r=1}^{K}\Pr\left[\left(N_{j_{r}}^{1},N_{j_{r}}^{2}\right)=\left(0,0\right)\right]\right\} \\
 & =2\left[\left(1-\phi\right)^{K}-\left(1-\phi-d_{12}\left(t\right)/2\right)^{K}\right]
\end{align*}
averaging over the choice of interactions and initial conditions,
the total distance obeys 
\begin{align*}
d_{12}\left(t+dt\right) & =2\sum_{K}P_{K}\left[\left(1-\phi\right)^{K}-\left(1-\phi-d_{12}\left(t\right)/2\right)^{K}\right]/S+\left(1-1/S\right)d_{12}\left(t\right)
\end{align*}
and using $dt=1/S$, $P_{K}=e^{-C}\frac{C^{K}}{K!}$ 
\begin{align*}
\frac{d}{dt}d_{12} & =2\sum_{K}e^{-C}\frac{C^{K}}{K!}\left[\left(1-\phi\right)^{K}-\left(1-\phi-d_{12}/2\right)^{K}\right]-d_{12}=2e^{-C\phi}\left[1-e^{-\frac{1}{2}Cd_{12}}\right]-d_{12}
\end{align*}
This has a fixed point at $d_{12}=0$ for any $C$. Using $\phi=W\left(C\right)/C$,
this fixed point loses stability at $C=e$, where a new stable fixed
point appears growing continuously from 0:
\begin{equation}
d_{12}^{*}=2\left[W\left(C\right)+W\left(-W^{2}\left(C\right)/C\right)\right]/C
\end{equation}
$d_{12}^{*}$ has a maximum at $C\approx7.14$, and tends to zero
as $C\rightarrow\infty$.

\section{\label{sec:non-zero beta proof}species with $N_{i}=1$ are isolated
when away from short cycle in the LF phase, when adding variability
and bi-directionality}

In the main text we show that for the minimal model (unidirectional
constant-strength interactions), for $\alpha>1$ and far away from
short cycles, the abundances are fixed at either $N_{i}=0,1$, with
the $N_{i}=1$ species separated from each other by extinct species.
Here we will prove that this also holds when there are positive bi-directional
interactions as well as bounded variability in the interaction strengths.

Consider a system where for any interaction with $\alpha_{ij}=\alpha>1$,
the reciprocal interaction is $\alpha_{ji}=\beta>0$. Adding variability
to these values will not change the proof, as long as it is the probability
distribution is bounded so that $\alpha_{ij}>1$ and $\alpha_{ji}>0$.
As the Erd\H{o}s-Rényi graph is tree-like almost everywhere and we
are interested in the behavior far away from the short cycles, we
need only show that there are no finite tree-like subgraphs that can
coexist at a fixed point, except for a single species in isolation.

Let us consider the LV equations on a finite connected tree with at
least two species, and assume by contradiction that it has a stable
fixed point where all species coexist with values $N_{i}^{*}>0$.
The variables $N_{i}$ can be rescaled so that the interactions are
symmetrized in the following manner \citep{pykh_lyapunov_2001}. Take
new variables $n_{i}=N_{i}/\gamma_{i}$, with $\gamma_{i}$ some constants
to be chosen later. The equation then gives a new LV system 
\begin{align}
\frac{dn_{i}}{dt} & =\gamma_{i}n_{i}\left(1/\gamma_{i}-n_{i}-\sum_{j}(\gamma_{j}/\gamma_{i})\alpha_{ij}n_{j}\right)\equiv\gamma_{i}n_{i}\left(1/\gamma_{i}-n_{i}-\sum_{j}a_{ij}n_{j}\right)\,,\label{eq:symmetrized LV on tree}
\end{align}
with interaction matrix $a_{ij}=(\gamma_{j}/\gamma_{i})\alpha_{ij}$.
The $\gamma_{i}$'s are chosen as follows: choose some species $i$
on the tree which interacts with species $j_{n}$, $n=1,..,K$, and
take $\gamma_{i}=1$. For each $j_{n}$, in order to have a symmetric
interaction $a_{ij_{n}}=a_{j_{n}i}$ one must set $\gamma_{j_{n}}=\gamma_{i}\sqrt{\frac{\alpha_{j_{n}i}}{\alpha_{ij_{n}}}}$.
As all $\alpha_{ij_{n}},\alpha_{j_{n}i}>0$, this choice of $\gamma_{j_{n}}$
is real and positive. This process is iterated for the species that
interact with the $j_{n}$, and so on until the values of $\gamma$
are set for the entire tree.

Now consider the fixed points of \ref{eq:symmetrized LV on tree},
$n_{i}^{*}=N_{i}^{*}/\gamma_{i}$. As $\gamma_{i}>0$, this is a feasible
fixed point with $n_{i}^{*}>0$, and from the assumption it is also
stable. This must be the only stable fixed point of the system, as
a symmetric LV system has a stable fixed point where all species coexist
iff it has a unique fixed point \citep{marcus_local_2022}.

We will now show in contradiction that another stable fixed point
can be constructed. Consider a new $\tilde{N}_{i}$ system, which
is the same as the original $N_{i}$ system but with $\beta=0$. Consider
all species that have no incoming interactions (there must be at least
one on a tree), which must have $\tilde{N}_{i}=1$ at a fixed point.
All outgoing interactions from these species must have $\tilde{N}_{i}=0$,
and continuing to move downstream all $\tilde{N}_{i}$ will be uniquely
determined to have either $\tilde{N}_{i}=0$ or $\tilde{N}_{i}=1$.
The fixed point found in this way is also a stable fixed point of
the $N_{i}$ system. This gives a fixed point with some $n_{i}=1/\gamma_{i}$
and some $n_{i}=0$, which is different from $n_{i}^{*}$, where for
all $i$, $n_{i}^{*}>0$, in contradiction to the assumption.
\end{document}